\documentclass[aps,prl,preprint,superscriptaddress]{revtex4-1}

\usepackage[pdftex]{graphicx}

\usepackage{color}

\begin{document}

\title{Nuclear structure beyond the neutron drip line: the lowest energy states in $^9$He via their T=5/2 isobaric analogs in $^9$Li}

\author{E. Uberseder}
\affiliation{Department of Physics \& Astronomy and Cyclotron Institute, Texas A\&M University, College Station, TX 77843, USA}

\author{G.V. Rogachev}
\affiliation{Department of Physics \& Astronomy and Cyclotron Institute, Texas A\&M University, College Station, TX 77843, USA}

\author{V.Z. Goldberg}
\affiliation{Department of Physics \& Astronomy and Cyclotron Institute, Texas A\&M University, College Station, TX 77843, USA}

\author{E. Koshchiy}
\affiliation{Department of Physics \& Astronomy and Cyclotron Institute, Texas A\&M University, College Station, TX 77843, USA}

\author{B.T. Roeder}
\affiliation{Department of Physics \& Astronomy and Cyclotron Institute, Texas A\&M University, College Station, TX 77843, USA}

\author{M. Alcorta}
\affiliation{TRIUMF, Vancouver, Canada}

\author{G. Chubarian}
\affiliation{Department of Physics \& Astronomy and Cyclotron Institute, Texas A\&M University, College Station, TX 77843, USA}

\author{B. Davids}
\affiliation{TRIUMF, Vancouver, Canada}

\author{C. Fu}
\affiliation{Shanghai Jiao Tong University, Shanghai, China}

\author{J. Hooker}
\affiliation{Department of Physics \& Astronomy and Cyclotron Institute, Texas A\&M University, College Station, TX 77843, USA}

\author{H. Jayatissa}
\affiliation{Department of Physics \& Astronomy and Cyclotron Institute, Texas A\&M University, College Station, TX 77843, USA}

\author{D. Melconian}
\affiliation{Department of Physics \& Astronomy and Cyclotron Institute, Texas A\&M University, College Station, TX 77843, USA}

\author{R.E. Tribble}
\affiliation{Department of Physics \& Astronomy and Cyclotron Institute, Texas A\&M University, College Station, TX 77843, USA}

\date{\today}

\begin{abstract}
The level structure of the very neutron rich and unbound $^9$He nucleus has been the subject of significant experimental and theoretical study.  Many recent works have claimed that the two lowest energy $^9$He states exist with spins $J^\pi=1/2^+$ and $J^\pi=1/2^-$ and widths on the order of hundreds of keV. These findings cannot be reconciled with our contemporary understanding of nuclear structure. The present work is the first high-resolution study with low statistical uncertainty of the relevant excitation energy range in the $^8$He$+n$ system, performed via a search for the T=5/2 isobaric analog states in $^9$Li populated through $^8$He+p elastic scattering.  The present data show no indication of any narrow structures. Instead, we find evidence for a broad $J^{\pi}=1/2^+$ state in $^9$He located approximately 3 MeV above the neutron decay threshold. 
\end{abstract}


\maketitle

The quest to understand the superheavy helium isotope $^9$He has been both long and fascinating.  Interest in $^9$He originates from its unusual ratio of neutron (N) to proton (Z) numbers (N/Z=3.5).  The largest N to Z ratio ($N/Z$=3) found among nucleon-bound isotopes belongs to the next heaviest helium isotope, $^8$He.  A rather unusual feature of  $^8$He is seen in its two-neutron separation energy, which is larger than in the less neutron rich isotope $^6$He. The isotope $^9$He, which is unstable to neutron decay, appears even more unusual. There has been significant experimental effort to determine the level structure of $^9$He.  A detailed history of $^9$He experimental studies has been recently given by \citet{NSR2013AL14}, and we will provide a brief overview of the current experimental status with respect to the ground and the first excited states in $^9$He, which are the main focus of this letter. 

The first observation of $^9$He via the $^9$Be($\pi^-$,$\pi^+$) reaction was reported in 1987 by \citet{NSR1987SE05} and its ground state was identified at 1.13$\pm$0.10 MeV above the neutron decay threshold.  \citet{NSR1987SE05} noted surprisingly good agreement between the energies of the peaks in the observed spectrum of $\pi^+$-mesons and the predictions of a shell model, attributing a $J^\pi=1/2^-$ spin assignment to the 1.13 MeV peak. Shortly thereafter, the $^9$He ground state was populated using the $^9$Be($^{13}$C,$^{13}$O) and $^9$Be($^{14}$C,$^{14}$O) reactions \cite{NSR1988BO20,NSR1995VO05,NSR1999BO26} and its energy was revised to 1.27$\pm$0.10 MeV. It appeared to be a narrow resonance with width of only 100$\pm$60 keV in the first high resolution study \cite{NSR1999BO26}. The narrow width of this state was in evident contradiction with the original expectations based on the shell model, and this fact precipitated the forthcoming experimental and theoretical studies.

With the advent of rare isotope beams, new experimental techniques to populate states in $^9$He have been explored. A study of the two-proton knock-out reaction from $^{11}$Be \cite{NSR2001CH31} was the first to identify a state with $\ell$=0 at an energy less than 0.2 MeV above the $^8$He+n threshold as the new ground state of $^9$He. It was declared that the 2s1/2-1p1/2 parity inversion first observed for the 7th neutron in neutron rich $^{11}$Be was also present in $^9$He. This unusual order of shell model states is usually explained by the residual interaction of the extra neutron with protons in the unfilled shell (see, for example, the classical work of \citet{NSR1960TA07}). Still, explanation of the parity inversion in $^9$He requires a better understanding of the $^8$He structure.

The majority of the experimental studies made after Ref. \cite{NSR1999BO26} supported the presence of a  narrow $J^\pi=1/2^-$ level at 1.3 MeV \cite{NSR2007FOZY,Falou2007,NSR2010JO06,NSR2013AL14}. The only investigation which argued against such a resonance was that of \citet{NSR2007GO24}, where the d($^8$He,p)$^9$He reaction was performed using $^8$He beam to populate states in $^9$He. However, the energy resolution ($\sim$0.8 MeV) in Ref.~\cite{NSR2007GO24} was inadequate to observe the $\sim$100 keV resonance directly. Also practically at the same time the opposite claim was made by \citet{NSR2007FOZY}, where the reaction d($^8$He,p)$^9$He was also used and the observation of a sharp structure at 1.3 MeV was inferred, albeit with statistics of just a few counts on top of a non-zero background.  There have been various model attempts (see history in Ref. \cite{NSR2013AL14}) to explain the narrow width of the $J^\pi=1/2^-$ state but the calculated widths have been 5-10 times larger than is found experimentally.  Additionally,  a recent {\it ab initio} calculation \cite{NSR2012NO09} supports the model conclusions.  
 
The existence of the $\ell$=0 resonance in $^9$He and its actual excitation energy has been a subject of much debate since it was first claimed in 2001 by \citet{NSR2001CH31}.  There is a huge variation of the results from near zero scattering length \cite{Falou2007,NSR2010JO06}, consistent with no or at most a very weak $\ell$=0 final state interaction, to scattering length -20 fm \cite{NSR2007GO24} corresponding to a strong resonance near the neutron decay threshold in $^9$He. Recently the d($^8$He,p)$^9$He reaction was studied again \cite{NSR2013AL14} at the SPIRAL facility. The observation of the ground 2s1/2 state close to the neutron decay threshold 0.18$\pm$0.085 MeV and 1p1/2 state at 1.2$\pm$0.1 MeV with a width in the range from 0 to 300 keV (with 130 keV giving the best fit) was reported. Moreover, the authors also obtained angular distributions, which supported the spin-parity assignments for the observed states as $J^\pi=1/2^+$ and $J^\pi=1/2^-$. Hence the latest experimental work again supports very unusual properties of the lowest states in $^9$He.  

In contrast, a very recent theoretical work connecting states in $^9$He to states in $^{10}$He \cite{For15} argues that the $J^\pi=1/2^+$ state cannot exist below 1.0 MeV (otherwise $^{10}$He must be neutron-bound) relative to the neutron threshold.

\begin{figure}
\includegraphics[scale=0.42]{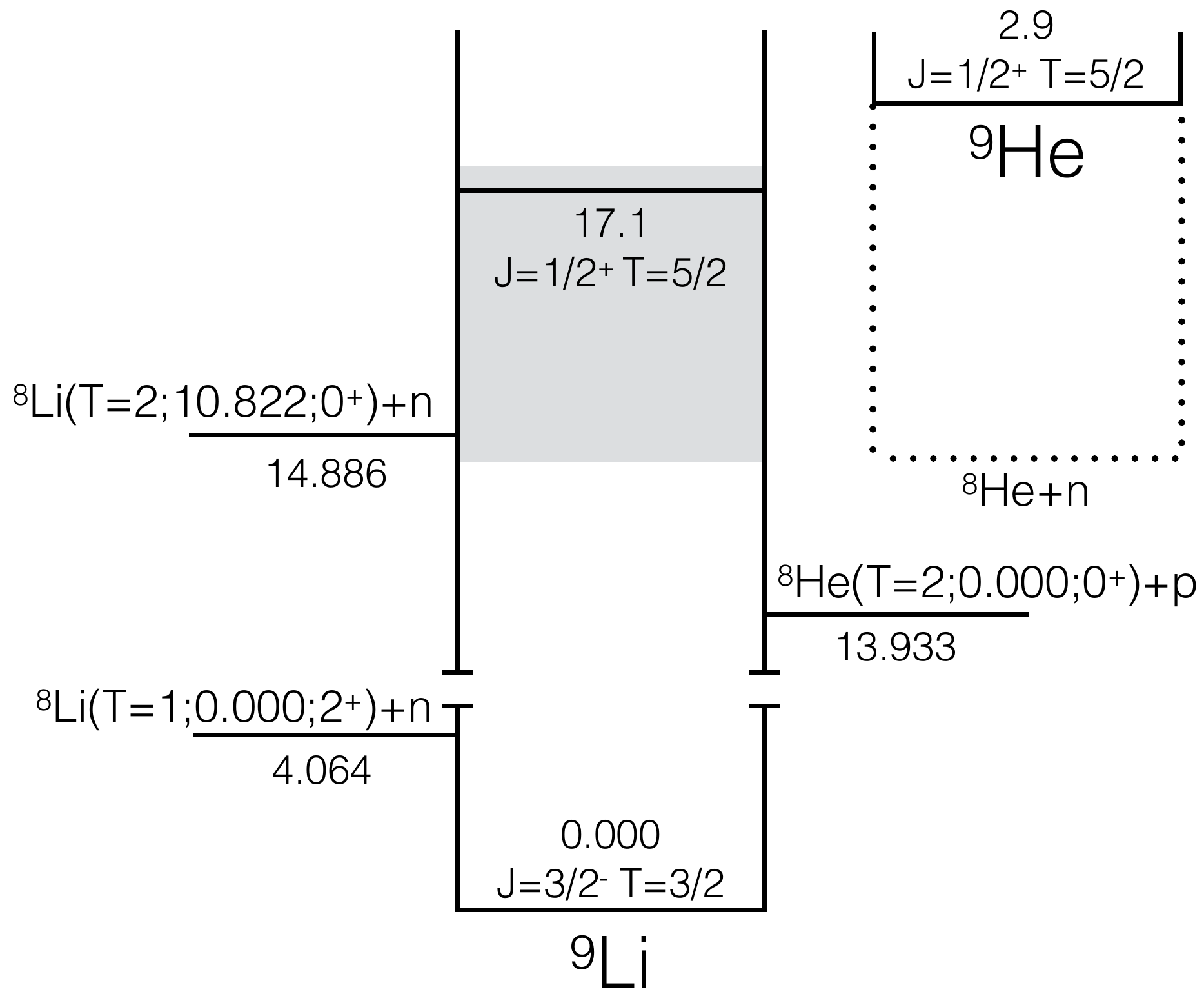}
\caption{\label{fig:levels} Level diagram indicating the excitation energy of $^9$Li probed in the current measurement (shaded region). The corresponding energies in $^9$He are shown for comparison.  All energies are in MeV.  The decay thresholds are calculated from Refs.~\cite{NSR2003AU03,NSR2004TI06}.}
\end{figure}

The exotic nature of $^{9}$He makes it a very difficult nucleus to probe experimentally, and previous studies have suffered from low counting statistics, poor energy resolution, or both. The present work constitutes a search for the low-lying $J^\pi=1/2^+$ and $J^\pi=1/2^-$ states in $^9$He via their T=5/2 isobaric analogs in $^{9}$Li populated directly through proton elastic scattering on $^{8}$He.  The reaction was performed with a $^{8}$He beam and utilized the thick target inverse kinematics (TTIK) method \cite{NSR1990AR24,NSR1993GO15,Gol98,NSR2010ROZV}, which has the advantage of measuring $^{8}$He$+p$ excitation functions for the elastic scattering with a single beam energy. In this technique, the incoming ions are slowed in the target gas (methane) and the recoil protons are detected from a scattering event. These recoil protons emerge from the interaction with $^8$He and hit Si detector array located at forward angles while the $^{8}$He ions are stopped in the gas, as the protons have smaller energy losses than the scattered ions. Due to straggling effects, the energy and angular spread of the incoming $^{8}$He ions increases as the ion traverses the scattering chamber. The spread of the beam in the chamber also depends upon its initial quality. Because we intended to populate the analog of the ground state in $^9$He that may be unbound by only 200 keV or less, we needed to reach a $^8$He+p center-of-momentum (CM) energy of about 1 MeV. Fig.~\ref{fig:levels} shows the corresponding neutron and proton thresholds in $^9$Li. The excellent quality of the reaccelerated $^8$He beam at the TRIUMF Isotope Separator and Accelerator (ISAC) facility, produced via the ISOL technique, enabled us to measure the $^{8}$He$+p$ elastic scattering cross section at much lower CM energies than has previously been possible~\cite{NSR2003RO07}. The horizontal dotted line in Fig.~\ref{fig:levels}  indicates the $^9$He neutron decay threshold with respect to the excitation energy of the T=5/2 isobaric analog states in $^9$Li. It is important to note that the only neutron decay allowed by isospin conservation for these states is to the T=2 excited state in $^8$Li (isobaric analog of the $^8$He ground state). In spite of its high excitation energy in $^8$Li, the T=2 state is very narrow (the upper limit is 12 keV);  all decays by nucleons are forbidden due to isospin conservation.  The shaded area in Fig.~\ref{fig:levels} represents the $^9$Li excitation energy region studied in this experiment and demonstrates that the isobaric analogs of the states in $^9$He that are barely unbound or even bound by few tens of keV would be populated in this experiment.  

The 32 MeV beam of $^{8}$He ions with  a $10^{4}$ pps average intensity impinged on a scattering chamber filled with  990 Torr of methane gas, entering the chamber through a thin (4 $\mu$m) Havar film. A windowless ionization chamber (IC) was installed close to the entrance window to count (for normalization) and identify the incoming ions. The $^8$He beam provided by ISAC was very pure; the only contaminant was $^8$Li$^{2+}$, at a level of 2\%, and was easily filtered using the IC. Three quadrant Si detectors (Micron Semiconductors MSQ25 type) were positioned symmetrically with respect to the beam axis at the distance of 513 mm from the entrance window, and provided information on the total energy of the recoil protons. A custom multi-anode position-sensitive proportional counter (MPPC) was installed in front of Si detectors to provide identification of the reaction products (using the $\Delta$E-E technique) as well as their transverse position. Detailed Monte Carlo studies of the present setup indicate CM energy resolutions for $^{8}$He$+p$ elastic scattering events ranging from 40 keV full-width half-maximum (FWHM) at CM energy of 3 MeV to 100 keV at the lowest energy of 0.8 MeV for the forward Si detector.

\begin{figure}
\includegraphics[scale=0.32]{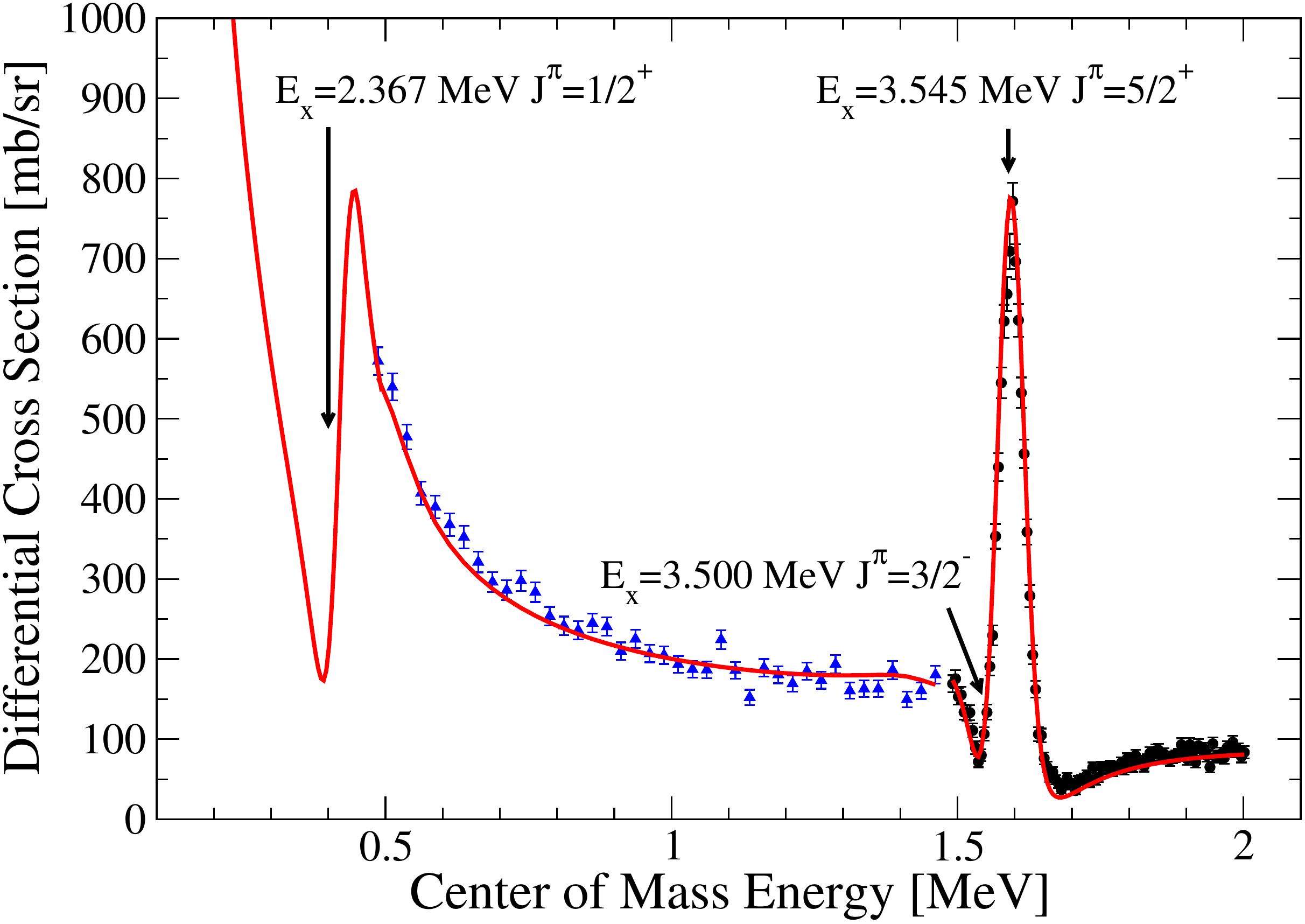}
\caption{\label{fig:c12pp} The $^{12}$C$+p$ elastic scattering cross section measured with the present setup at TRIUMF (blue triangles) and TAMU (black circles).  The red line is the R-matrix calculation (see text for the details).}
\end{figure}

Measurements with $^{12}$C beam were performed to test the experimental setup and to verify the analysis procedures. Fig.~\ref{fig:c12pp} shows the spectrum of protons from $^{12}$C+p elastic scattering measured in two different runs. The lower energy data (blue triangles) were measured at the Cyclotron Institute at Texas A\&M University (TAMU), and the higher energy data (black circles) were measured at the TRIUMF ISAC facility just before the $^8$He main production run. The experimental setup was identical in all measurements. The red curve is an R-matrix calculation (not a fit), convoluted with experimental energy resolution. Parameters for the R-matrix calculations were obtained by fitting the differential cross sections of \citet{NSR1976ME22}. The agreement reflects the reliability of the analysis procedures.

\begin{figure}
\includegraphics[scale=0.62]{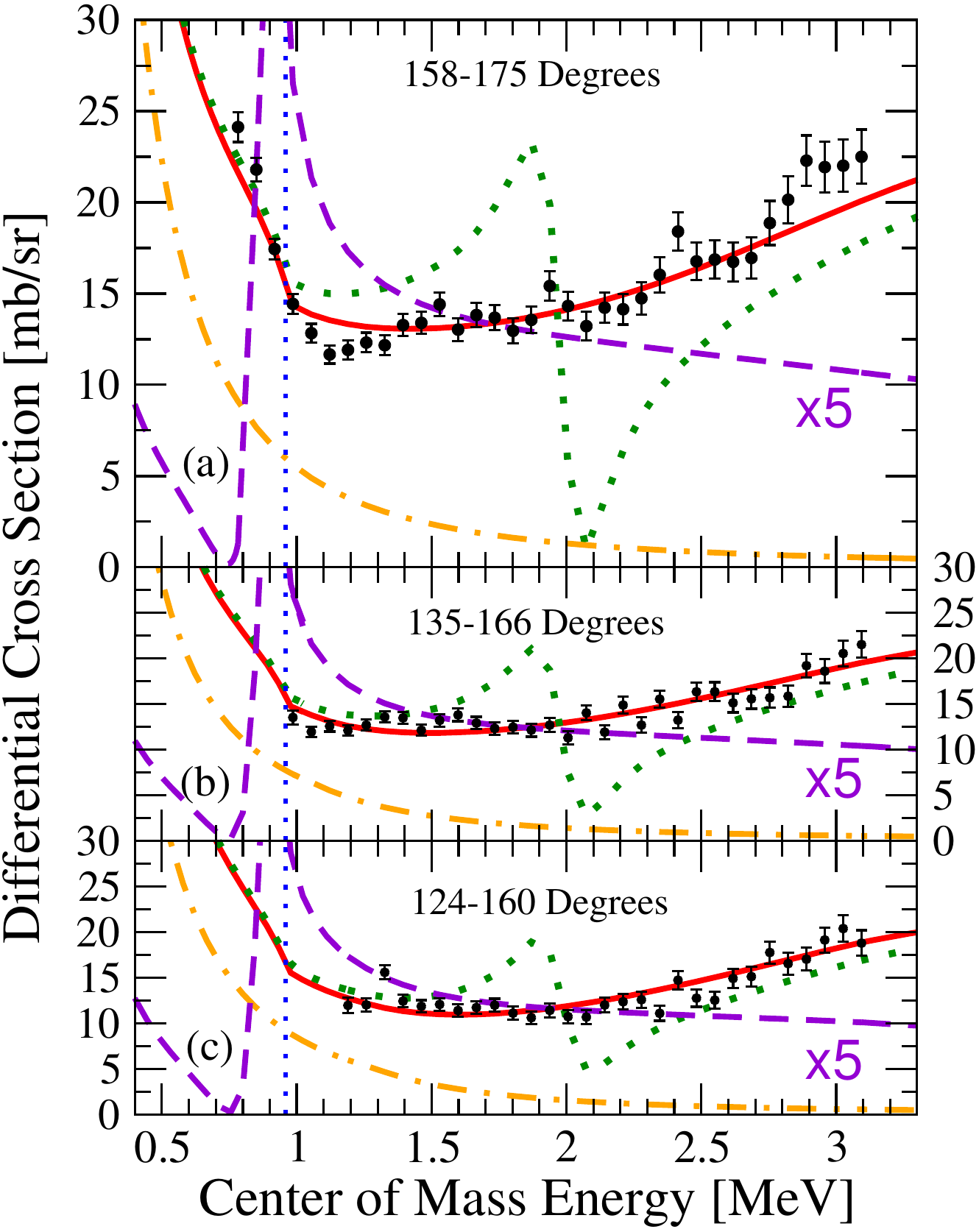}
\caption{\label{fig:he8pp} $^8$He$+p$ elastic scattering excitation functions measured at three different lab. angles. The corresponding CM scattering angles are functions of energy with range shown for each section.
The red solid curve is the best R-matrix fit. The orange dash-dotted curve is the Rutherford cross section. The green dotted curve demonstrates the sensitivity of these data to the hypothetical narrow T=5/2 1/2$^-$ state in $^9$Li. The purple dashed line shows the effect of a narrow T=5/2 1/2$^+$ state in $^9$Li (curve has been divided by 5 to appear on scale).  The $^8$Li(T=2;$E_x = 10.822$ MeV; 0$^+$)+n threshold is shown as a dotted blue line.}
\end{figure}

Fig.~\ref{fig:he8pp} shows the excitation functions for $^{8}$He$+p$ elastic scattering obtained in the present measurement. The error bars indicate the statistical uncertainty. The individual spectra, from top to bottom, correspond to proton detection in the central detector, the inner halves of the outer detectors, and the outer halves of the outer detectors, respectively. Scattering events of varying energies take place at different distances from the detectors, and therefore at different laboratory angles. The corresponding CM angles are shown for each spectrum in Fig.~\ref{fig:he8pp}. The protons emitted with low energies at higher angle must traverse a longer path before reaching a Si detector and thus have greater energy loss to the gas.  As such, the lower detection limit in CM energy increases from the top to the bottom plot.

\begin{figure}
\includegraphics[scale=0.35]{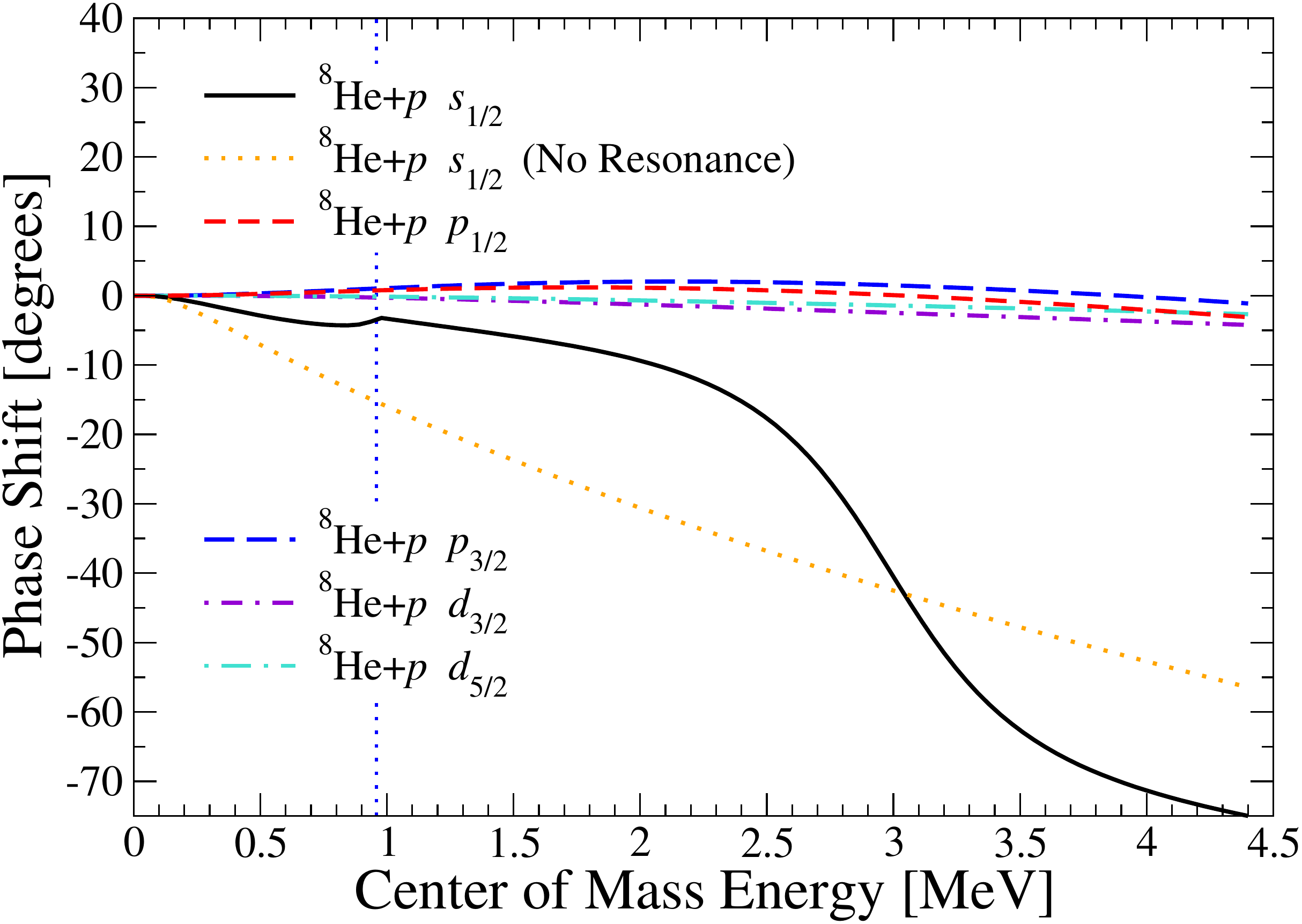}
\caption{\label{fig:phase} The $^8$He$+p$ phase shifts for the various partial waves determined from the R-matrix fit to the $^8$He$+p$ excitation functions. All but the s-wave ({\it s}$_{1/2}$) phase shift, shown as the black solid curve, are featureless and close to zero at the measured energies. The pure potential model phase shift that does not include the broad T=5/2 1/2$^+$ resonance is shown as the orange dotted curve. The $^8$Li(T=2;$E_x = 10.822$ MeV; 0$^+$)+n threshold is shown as a dotted blue line.}
\end{figure}

The spectra in Fig.~\ref{fig:he8pp} are rather featureless with the exception of a dramatic rise of the cross section at an energy corresponding to the $^8$Li(T=2;$E_x = 10.822$ MeV; 0$^+$)+n threshold, as seen in Fig.~\ref{fig:he8pp}a. As demonstrated in Fig.~\ref{fig:he8pp}, this rise cannot be explained by Rutherford scattering. The T=3/2 levels in $^9$Li in this excitation region are unknown, therefore a hybrid R-matrix approach based on the ideas of Refs.~\cite{Rob65,TAR68} was utilized in the analysis. In this approach the effect of the unknown T=3/2 levels, which decay to many isospin-allowed open channels, is described by an optical model potential (details of the analysis will be published elsewhere). The introduction of the optical model increases the parameter space, though fortunately the  $\ell$=0 partial wave dominates the excitation function in the measured energy region ($\it kR$$\sim$1), as expected from the nearly isotropic angular distributions.  Contributions to the cross section from other partial waves were found to be negligible (see Fig.~\ref{fig:phase}).

The narrow $J^\pi=1/2^-$ resonance, suggested to be 1.3 MeV above the neutron decay threshold in $^9$He \cite{NSR1999BO26,NSR2013AL14}, would have been easily observed in our data at an energy of about 1.2 MeV above the $^8$Li(0$^+$,T=2)+n threshold of 14.886 MeV. Instead, the excitation function in that energy region is featureless at all angles (Fig.~\ref{fig:he8pp}).  The manifestation of the $J^\pi=1/2^-$ resonance with a 100 keV width in the energy region of interest is shown in Fig.~\ref{fig:he8pp} with green dotted curve. This calculation properly treated the neutron decay of the resonance to the $^8$Li(T=2;$E_x = 10.822$ MeV;$0^+$), which is the dominant decay channel given that the neutron to proton ($^8$He$+p$) reduced width amplitude ratio is fixed by the isospin Clebsh-Gordon coefficients ($\gamma_n$/$\gamma_p$=2). The experimental resolution is 50 keV at this CM energy and was also taken into account in the R-matrix calculation. To escape observation, the state would need to be as narrow as 20 keV in $^9$He, i.e. even narrower than it was claimed in previous measurements. Another possible way this state could remain unobserved in the present measurement would be if it was strongly isospin impure. In this case the decays to many open T=1 channels would make the resonance broader and weaker.  Our calculations show that the isospin mixing would have to be nearly 50\% to make this possible.  Isospin mixing on such a scale would need a special explanation.
 
As stated above, the most natural explanation of the cross section rise near the $^8$Li(T=2;$E_x = 10.822$ MeV;$0^+$)+n decay threshold is a manifestation of the Wigner cusp \cite{Wigner1948}.  This decay threshold, located at an excitation energy of 14.9 MeV in $^9$Li, is significant only for the T=5/2 resonances. Closing of this channel leaves $^8$He$+p$ as the only open isospin-allowed decay channel for the T=5/2 resonances and the cross section rises dramatically to preserve the incoming particle flux. To reproduce the threshold effect in question, a broad T=5/2 $J^\pi=1/2^+$ resonance needed to be introduced, with a width comparable to the distance between the resonance excitation energy and the $^8$Li(T=2;$E_x = 10.822$ MeV;$0^+$)+n threshold. The actual parameters of the T=5/2 $J^\pi=1/2^+$ resonance are fairly sensitive to the shape of the observed cusp. The best fit (shown as a solid red curve in Fig.~\ref{fig:he8pp}) is achieved with the $\gamma_p$=0.5 MeV$^{1/2}$ and 17.1 MeV excitation energy for this state. The resulting s-wave phase shift, shown with the black solid curve in Fig.~\ref{fig:phase}, clearly demonstrates the influence of the broad s-wave resonance on the behavior of the phase shift. It produces the sudden change of the phase shift derivative near the $^8$Li(T=2;$E_x = 10.822$ MeV; 0$^+$)+n decay threshold that is in turn responsible for the observed rise of the cross section.  A low energy resonance with properties claimed in Ref. \cite{NSR2013AL14} is incompatible with the measured excitation function (see Fig.~\ref{fig:he8pp}), as it leads to dramatic effects near the $^8$Li(T=2;$E_x = 10.822$ MeV; 0$^+$)+n decay threshold. The cross section would be five times higher than observed and would have a very distinct shape that is different from the experimental data.
  
Taking into account the shift functions of the $^8$He$+p$ and $^8$He$+n$ systems, the T=5/2 $J^\pi=1/2^+$ state physically appears in $^9$He at CM energy of $\sim$3 MeV above the neutron decay threshold with a width of $\sim4$ MeV.

In summary, we report the first high resolution search with low statistical uncertainty for low-lying states in $^9$He through their T=5/2 isobaric analogs in $^{9}$Li. We did not observe any narrow structures within the energy range of interest, and ruled out an existence of a narrow $J^\pi=1/2^-$ state in $^9$He.  We provided  evidence for a very broad T=5/2 state with spin $J^{\pi}=1/2^+$ at an excitation energy of 17.1 MeV in $^9$Li. This corresponds to a  virtual broad ($\sim$4 MeV) state in $^9$He at 3 MeV energy above the neutron decay threshold. Two long-standing problems are resolved by these results. First, the mysterious discrepancy by a factor of 5-10 between the theoretical predictions and the experiment for the width of the low lying $J^\pi=1/2^-$ state in $^9$He has been eliminated by showing that there are no narrow resonances in $^9$He at energies between 0 and 2.2 MeV above the neutron decay threshold. Second, it was shown that the actual energy of the $J^\pi=1/2^+$ state in $^9$He is far above that determined in Ref.~\cite{NSR2001CH31} and more recently in Ref.~\cite{NSR2013AL14} and that it has to be a very broad state. Search for the states in neutron rich nuclei through their isobaric analogs has been proven to be very powerful technique even for the very difficult case of $^9$He (all states are unbound and very broad). We expect that this approach will quickly gain popularity and will become an important tool for nuclear structure studies of very neutron rich isotopes with rare isotope beams.

\begin{acknowledgments}
The authors are very grateful to the accelerator physicists and technical staff at the TRIUMF facility for their excellent and exceptionally professional work and to the management of the TRIUMF laboratory for providing ideal environment for successful experiment. The authors acknowledge that this material is based upon their work supported by the U.S. Department of Energy, Office of Science, Office of Nuclear Science, under Award No. DE-FG02-93ER40773. The author G.V.R. and H.J. are also supported by the Welch Foundation (Grant No. A-1853).   
\end{acknowledgments}

\end{document}